\documentclass[aps,prd,twocolumn,floatfix,noshowpacs,tightenlines,noshowkeys,superscriptaddress,amsmath,amssymb,nofootinbib]{revtex4}
\usepackage{amssymb,amsbsy,epsfig,color,graphicx}
\usepackage{color}
\usepackage{[longtable}
\usepackage{array}
\usepackage{dcolumn}   % needed for some tables
\usepackage{cellspace}
\usepackage{mathtools}
\usepackage{amstext}
\usepackage{amssymb}
\usepackage{stmaryrd}
\usepackage{stackrel}
\usepackage{graphicx}
\usepackage{esint}
\usepackage[utf8]{inputenc}
\usepackage{blindtext}
\usepackage{float}
\restylefloat{table}
\usepackage{booktabs}
\usepackage{enumitem} 

\usepackage{etoolbox} % for \appto
\usepackage{lipsum} % for mock text
\usepackage[capitalize]{cleveref}
\usepackage{multirow}
\usepackage[caption=false]{subfig}
\renewcommand\[{\begin{equation}}
\renewcommand\]{\end{equation}}

\newcommand{\ba}{\begin{eqnarray}}
\newcommand{\ea}{\end{eqnarray}}

\makeatletter

\appto{\appendix}{%
\@ifstar{\def\theequation@prefix{A.}}%
{}%
}
\makeatother

\begin{document}

\title{Transmutation of nonlocal scale in infinite derivative field theories}

\author{Luca Buoninfante}
\affiliation{Dipartimento di Fisica "E.R. Caianiello", Universit\`a di Salerno, I-84084 Fisciano (SA), Italy}
\affiliation{INFN - Sezione di Napoli, Gruppo collegato di Salerno, I-84084 Fisciano (SA), Italy}
\affiliation{Van Swinderen Institute, University of Groningen, 9747 AG, Groningen, The Netherlands}

\author{Anish Ghoshal}
\affiliation{Laboratori Nazionale di Frascati-INFN, C.P. 13, 100044, Frascati, Italy}
\affiliation{Dipartimento di Matematica e Fisica,  Universit\`a Roma Tre, 00146 Rome, Italy}

\author{Gaetano Lambiase}
\affiliation{Dipartimento di Fisica "E.R. Caianiello", Universit\`a di Salerno, I-84084 Fisciano (SA), Italy}
\affiliation{INFN - Sezione di Napoli, Gruppo collegato di Salerno, I-84084 Fisciano (SA), Italy}

\author{Anupam Mazumdar}
\affiliation{Van Swinderen Institute, University of Groningen, 9747 AG, Groningen, The Netherlands}

%	\date{\today}

\begin{abstract}
In this paper we will show an ultraviolet -infrared  connection for ghost-free infinite derivative field theories where the Lagrangians are made up of exponentials of entire functions. In particular, for $N$-point amplitudes a new scale emerges in the infrared from the ultraviolet, i.e. $M_{\rm eff}\sim M_s/N^\alpha,$ where $M_s$ is the fundamental scale beyond the Standard Model, and $\alpha>0$  depends on the specific choice of an entire function and on whether we consider zero or nonzero external momenta.  We will illustrate this by first considering a scalar toy model with a cubic interaction, and subsequently a scalar toy-model inspired by ghost-free infinite derivative theories of gravity. We will briefly discuss some phenomenological implications, such as making the nonlocal region macroscopic in the infrared.
\end{abstract}

\maketitle

%%%%%%%%%%%%%%%%%%%%%%%%%%%%%%%%%%%%%%%%%%%%%%%%%%%%%%%%%%%%%%%%%%%%%%%%

\section{Introduction}\label{intro}

It has been known that infinite derivative field theories can give rise to nonlocal physics, which has been studied extensively 
in~\cite{Yukawa:1950eq,efimov,Krasnikov,Kuzmin,Moffat,Tomboulis:1997gg,Biswas:2014yia,Tomboulis:2015gfa,Ghoshal:2017egr,Buoninfante:2018mre},  in the context of string field theory~\cite{Witten:1985cc,Freund:1987kt,eliezer}, and in the context of gravity~\cite{Tseytlin:1995uq,Siegel:2003vt,Biswas:2005qr,Biswas:2011ar,Modesto:2011kw,Biswas:2016etb}. The main properties of such theories can be captured by form-factors, which are not polynomials in the derivatives, but   {\it transcendental entire functions}, which ensures the ghost-free condition at the perturbative level, see \cite{Biswas:2011ar,Biswas:2013kla,Biswas:2016etb,sen-epsilon,carone,chin,Briscese:2018oyx}. Since, an entire function is defined as a function which has no poles in the complex plane, no new degrees of freedom (d.o.f) arise other than the standard local d.o.f. Typically, such theories have smooth infrared (IR) limit from the ultraviolet (UV) below the scale of nonlocality given by $M_{s}\leq M_{p},$ where $M_{p}=1.2\times 10^{19}$~GeV is the $4$ dimensional Planck mass. 

 It has been earlier shown that presence of such form-factors could improve the UV behavior of the theory and this has stimulated a
 deeper investigation of these models from both a physical and a mathematical point of view \cite{efimov}. In 
 Refs.~\cite{Krasnikov,Kuzmin,Tomboulis:1997gg}, the authors have studied nonlocality in the context of gauge theories and ghost-free gravity in 4 
 dimensions~\cite{Biswas:2005qr,Biswas:2011ar,Modesto:2011kw,Biswas:2013kla,Talaganis:2014ida} around Minkowski spacetime, and in (A)dS background~\cite{Biswas:2016etb}. Recently, the three dimensional version of ghost-free, infinite derivative theory of gravity (IDG) has been constructed~\cite{Mazumdar:2018xjz}.

In the context of gravity, the propagator is suppressed by the exponential of an entire function in order not to introduce any new dynamical d.o.f other than the massless graviton as in the Einstein general relativity. From a classical point of view, the presence of such form-factors can  improve the short-distance behavior of the theory by resolving blackhole \cite{Biswas:2011ar,Biswas:2013cha,Biswas:2016etb,Edholm:2016hbt,Frolov,Frolov:2015bia,Frolov:2015usa,Buoninfante,Koshelev:2017bxd,Buoninfante:2018xiw,Koshelev:2018hpt,Buoninfante:2018rlq,Buoninfante:2018stt,Giacchini:2018wlf,Buoninfante:2018xif}, extended objects~\cite{Boos} and cosmological singularities \cite{Biswas:2005qr,Biswas:2010zk,Biswas:2012bp,Koshelev:2012qn,Koshelev:2018rau}. While, from a quantum point of view, it is believed that nonlocality can improve the UV behavior of the theory \cite{Modesto:2011kw,Talaganis:2014ida,Biswas:2014yia}. 

In Ref.~\cite{Ghoshal:2017egr} it was shown that the Abelian Higgs potential is free from instabilities as the $\beta$-function vanishes for energies above the nonlocal scale, $p^2>M_s^2$, and nonlocal extensions of finite gauge theories have been studied in Refs.~\cite{Moffat,Tomboulis:1997gg}. In Refs.~\cite{Biswas:2014yia,Talaganis:2016ovm} the authors have shown that the $2\leftrightarrow 2$ scattering amplitude can be exponentially suppressed above the nonlocal scale. It was also shown that at finite temperatures these nonlocal theories exhibit properties very similar to the Hagedorn gas~\cite{Biswas:2009nx,Biswas:2010xq,Biswas:2010yx}, especially for a p-adic type action. In the cosmological context such nonlocal theories have shown interesting possibility for explaining cosmic inflation~\cite{inflation}.

The aim of this paper is to show that a new scale emerges in ghost-free infinite derivative field theories in the IR. We will illustrate this in simple scalar toy-models by computing $N$-point amplitudes, and understanding their behavior for a large number of external legs, $N\gg 1.$ We will consider both non-zero and zero external momenta. In the former case, we will be in the physical scenario of scattering amplitudes, while the latter case can be applied to study the interaction among the constituents of bound-systems, like condensates, which can be seen as made up of off-shell quanta. We will show that the larger is the number of particles participating in the interaction process, the more exponentially suppressed will be the amplitude. Such a phenomena can be also interpreted as if the nonlocal scale smoothly shifts as a function of $N,$  $M_s\longrightarrow  M_s/N^{\alpha},$ where $\alpha>0$ depends on the form of the entire function, which we will discuss below.  This feature highlights an intriguing connection between UV and IR, such that the length scale of nonlocality can swell up to larger scales, as $M_{s}^{-1} \longrightarrow N^{\alpha}M_{s}^{-1}$, for $N\gg 1.$ 

Such a scaling behaviour also happens in string theory, i.e. in the context of fuzz-ball, where the compact gravitational system made up of branes and strings can swell up to scales larger than the Schwarzschild radius~\cite{Mathur:2005zp}. In fact, such a swelling effect in the case of nonlocality has already been postulated from entropic arguments in the context of gravity - in order to maintain the {\it Area}-Law of gravitational entropy, where the effective scale of nonlocality shifts to: $M_{\rm eff}\sim M_s/\sqrt{N}$, see~\cite{Koshelev:2017bxd}. Here we will obtain such a scaling relationship via scattering scattering amplitudes.

In Section \ref{nlft}, we will introduce infinite derivative field theories. In Section \ref{scal-cub}, as a warm up exercise, we will compute $N$-point amplitudes for a massless scalar field with a nonlocal kinetic term and a standard cubic interaction. In Section \ref{toy-model-gravity}, we will discuss $N$-point amplitudes in a scalar toy-model inspired by IDG, which can mimic the graviton self-interaction up to cubic order in the metric perturbation around the Minkowski background. In Section \ref{concl}, we will discuss our results and present the conclusions.

%%%%%%%%%%%%%%%%%%%%%%%%%%%%%%%%%%%%%%%%%%%%%%%%%%%%%%%%%%

\section{Infinite derivative field theory}\label{nlft}

Let us consider a simple model of self-interacting scalar field \cite{Tomboulis:2015gfa,Buoninfante:2018mre}~\footnote{Throughout the paper we will use the mostly positive metric convention, $(-+++),$ and work with Natural Units, $\hbar=1=c$. }
\begin{equation}
S=\frac{1}{2}\int d^4x \phi(x)e^{f(\Box_s)}(\Box-m^2)\phi(x)-\int d^4x V\left(\phi(x)\right),\label{non local-action}
\end{equation}
where $f(\Box_s)$ is an entire function, $\Box_s\equiv \Box/M_s$ with $\Box$ being the flat d'Alembertian and $M_s$ is the scale of nonlocality at which new physics should manifest, $m$ is the mass of the field $\phi(x)$ and $V(\phi(x))$ is the potential whose functional form can be either local or nonlocal, as we will see below. Note that the exponential form-factor in Eq.\eqref{non local-action} can be moved from the kinetic to the interaction term by making the following field redefinition: $\tilde{\phi}=e^{\frac{1}{2}f(\Box_s)}\phi$. From this last observation, it is clear that nonlocality plays a crucial role {\it only} when the interaction is switched on \cite{Buoninfante:2018mre}. However, below the cut-off scale $M_s$ the theory smoothly interpolates to a local theory, recovering all its predictions~\footnote{Remarkably, exponential form-factors of the kind $(e^{\alpha'\Box}\phi)^3,$ also appears in string field theory \cite{Witten:1985cc,eliezer,Tseytlin:1995uq,Siegel:2003vt} and p-adic strings \cite{Freund:1987kt}, where $M_s$ is related to the tension of the string, $\alpha'\sim 1/M_s^2$.}.

As discussed in Refs.~\cite{Tomboulis:1997gg,Buoninfante:2018mre}, in infinite derivative field theories, amplitudes and correlators are well-defined in the Euclidean signature for momenta $p^2\geq M_s^2$. All the amplitudes need to be defined in the Euclidean space from the beginning. Also from a physical point of view, the Minkowski signature is not a sensible choice beyond the nonlocal scale, as for $\Box\geq M_s^{2}$,  causality is violated. However, there is nothing which prohibits probing such a system with a large number, $N,$ of on-shell states with  $\Box\ll M_s^2$. In this case we need to compare $N^{\alpha} p^2$ with the cut-off $M_s$, where $\alpha$ depends on the choice of $f(\Box_{s})$, see the discussions below. Furthermore, once the propagator and the vertices are dressed  by including all quantum corrections, 
there are {\it no} divergences emerge in $s,~t,~u$ channels~\cite{Talaganis:2016ovm,Buoninfante:2018mre}. We will show this explicitly here as well.

%%%%%%%%%%%%%%%%%%%%%%%%%%%%%%%%%%%%%%%%%%%%%%%

\section{Scalar field with cubic vertex interaction }\label{scal-cub}

As a warm up exercise, let us consider a simple toy model of infinite derivative massless scalar field with cubic interaction and form-factor $e^{f(\Box_s)}=e^{(-\Box/M_s)^n}.$ The corresponding Euclidean action reads
\begin{equation}\label{cubic-action}
S=\int d^4x \left(\frac{1}{2}\phi(x)e^{(-\Box_s/M_s^2)^n}\Box\phi(x) + \frac{\lambda}{3!}\phi^3(x)\right),
\end{equation}
where $\lambda$ being the coupling constant, and the Euclidean bare propagator is given by
\begin{equation}
\Pi(p)=\frac{e^{-(p^2/M_s^2)^n}}{p^2},\label{euclid-propag}
\end{equation}
which is exponentially suppressed in the UV regime, $p^2\gg M_s^2$, with $p^2 =(p^4)^2+\vec{p}^2> 0$ being the squared of the $4$-momentum in an Euclidean signature $p\equiv (ip^4,\vec{p})$. The bare vertex is just a constant:
\begin{equation}
V(k_1,k_2,k_3)=\lambda.\label{euclid-vertex}
\end{equation}
%

%%%%%%%%%%%%%%%%%%%%%%%%%%%%%%%%%%%%%%%

\subsubsection{Dressing the propagators}

As in local quantum field theory, the dressed propagator takes into account of all possible infinite quantum corrections coming from higher loop contributions. For the action in Eq.(\ref{cubic-action}), it is given by \cite{Talaganis:2014ida,Biswas:2014yia,Buoninfante:2018mre}~\footnote{The dressed propagator in Eq.\eqref{dressed} has a more complicated pole structure compared to the bare one. Indeed, the equation $p^2+\Sigma(p)e^{-(p^2/M_s^2)^n}=0,$ can have real solutions and an infinite number of complex conjugate solutions. The $1$-loop dressed propagator for the action in Eq.\eqref{cubic-action}, besides infinite complex conjugate poles, shows also the presence of a real ghost-mode, which may cause instabilities \cite{Buoninfante:2018mre}. However, this feature is model-dependent, indeed for the gravitational toy-model in Section \ref{toy-model-gravity} the dressed propagator has a massless pole, $p^2=0,$ plus a stable tachyon-mode besides infinite complex conjugate poles, and no ghosts \cite{Talaganis:2014ida}.\label{foot-poles}}:
\begin{equation}
\Pi_{\rm dress}(p)=\frac{e^{-(p^2/M_s^2)^n}}{p^2+\Sigma(p)e^{-(p^2/M_s^2)^n}},\label{dressed}
\end{equation}
where the self-energy is defined as
\begin{equation}
\Sigma(p)=\lambda^2\int \frac{d^4k}{(2\pi)^4} \frac{e^{-(k^2/M_s^2)^n}e^{-((k-p)^2/M_s^2)^n}}{k^2(k-p)^2}. \label{self}
\end{equation}
Let us examine the simpler case $n=1$, for which we can explicitly give analytic results for  the $1$-loop self-energy contribution~\cite{Buoninfante:2018mre}
\begin{equation}
\begin{array}{rl}
\Sigma^{(1)}(p)= & \displaystyle \frac{\lambda^2}{16\pi^2}\left[\frac{2M_s^2}{p^2}\left(e^{-p^2/2M_s^2}-e^{-p^2/M_s^2}\right)\right.\\
&\displaystyle \left.+{\rm Ei}\left(-\frac{p^2}{2M_s^2}\right)-{\rm Ei}\left(-\frac{p^2}{M_s^2}\right)\right]\,,\label{1-loop self-energy}
\end{array}
\end{equation}
where $${\rm Ei}(x)=\int\limits^{x}_{-\infty}dt\,\frac{e^{t}}{t},$$ is the so called {\it exponential integral function}.
The exact expression for the self-energy at $1$-loop in Eq.\eqref{1-loop self-energy} is quite complicated, however  in the regime where the exponential form-factors are important, $p^2\gg M_s^2,$ one can show that the self-energy goes like $e^{-p^2/M_s^2}$ and the dressed propagator behaves like~\cite{Buoninfante:2018mre}
\begin{equation}
\Pi_{\rm dress}(p)=\frac{e^{-p^2/M_s^2}}{p^2+e^{-p^2/M_s^2}\cdot e^{-p^2/M_s^2}}~~\xrightarrow{UV}~~\frac{e^{-p^2/M_s^2}}{p^2}.\label{dressed-uv}
\end{equation}
We would expect a similar scenario to hold for any powers of $\Box^{n}$, for $n >0$. From Eq.\eqref{dressed-uv}, it is clear that for this model the bare and dressed Euclidean propagators have the same UV behavior, see Eq.(\ref{euclid-propag}). However, this is model dependent, and this will not be guaranteed for other examples, such as the one we would consider in section \ref{toy-model-gravity}.

%%%%%%%%%%%%%%%%%%%%%%%%%%%%%%%%%%%%%%%%%%%%%%%%%

\subsubsection{Dressing the vertices}

The dressed vertex at $1$-loop is defined by replacing the bare vertex with a {\it triangle} made up of three bare vertices and three internal propagators:
\begin{equation}
V_{\rm dress}(k_1,k_2,k_3)=\lambda^3\int \frac{d^4k}{(2\pi)^4}\Pi(k_1)\Pi(k_2)\Pi(k_3).\label{dressed-vertex}
\end{equation}

In particular, in the UV region, $k_{i}^{2}> M_{s}^{2}$, the dressed vertex in Eq.\eqref{dressed-vertex} can be computed as follows:
\begin{equation}
V_{\rm dress}(k_1,k_2,k_3)~~\xrightarrow{UV}~~\lambda^3e^{-\frac{1}{3}k^2_1/M_{s}^{2}}e^{-\frac{1}{3}k^2_2/M_{s}^{2}}e^{-\frac{1}{3}k^2_3/M_{s}^{2}}\,.
\label{dressed-vertex-uv}
\end{equation}
From Eq.\eqref{dressed-vertex-uv}, we note that the dressed vertex is not a constant anymore, but it has acquired an exponentially suppressed behavior in the high energy regime.

%%%%%%%%%%%%%%%%%%%%%%%%%%%%%%%%%%%%%%%%%%%%%%%%%%%%%%%%%%%%%%%%%%%
\subsection{$N$-point scattering amplitude}
\begin{figure}[t]
	\begin{center}
		\includegraphics[scale=0.52]{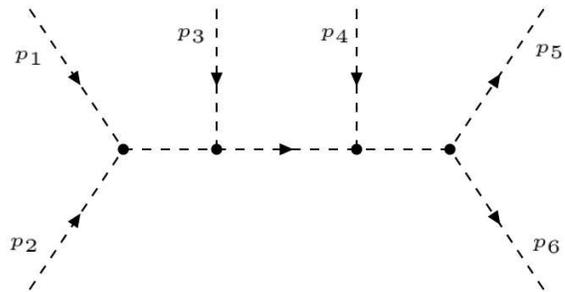}
		\caption{$6$-point amplitude for scalar cubic interaction. The blobs correspond to dressed vertices.}
	\end{center}\label{Fig_1}
\end{figure}
We now wish to compute $N$-point amplitudes, $\mathcal{M}_N,$ for the action in Eq.\eqref{cubic-action}. 
Let us consider $n=1$ to start with, and then we will generalize to generic powers of $\Box$. A generic tree-level $N$-point amplitude for the action in Eq.\eqref{cubic-action} will be made of $N$ external legs, $N-2$ vertices and $N-3$ internal propagators; see Fig. $1$.
The simplest scattering amplitude we can construct is a $4$-point diagram, whose behavior in the UV regime is given by:
\begin{equation}
\mathcal{M}_4~~\xrightarrow{UV}~~\lambda^6\frac{e^{-\frac{5}{3}p^2/M_s^2}}{p^2},\label{4-point-cubic}
\end{equation}
with $p$ being the sum of the two ingoing (or, equivalently, outgoing) momenta, $p_1+p_2=p_3+p_4\equiv p.$ Similarly, by increasing the number of external legs to $6$, we can consider a $6$-point amplitude as in Fig. $1$~\footnote{In principle, we can also consider more complicated diagrams, but in the large $N$ limit all the correct features related to the exponential form-factors can be captured universally.}, where our convention is that all $p_i$ with $i=1,2,3,4$ are ingoing, while $p_5$ and $p_6$ are outgoing momenta. From the conservation law of the total $4$-momentum, we have:
\begin{equation}
p_1+p_2+p_3+p_4=p_5+p_6.\label{cons-law-6}
\end{equation}
The $6$-point amplitude in the UV regime then reads:
\begin{equation}
\!\!\mathcal{M}_6\!\xrightarrow{UV}\!\lambda^{12}\!
\frac{e^{-\frac{5}{3}\frac{(p_1+p_2)^2}{M_s^2}}e^{-\frac{5}{3}\frac{(p_1+p_2+p_3)^2}{M_s^2}}e^{-\frac{5}{3}\frac{(p_1+p_2+p_3+p_4)^2}{M_s^2}}}{(p_1+p_2)^2(p_1+p_2+p_3)^2(p_1+p_2+p_3+p_4)^2}.
\label{6-point-cubic}
\end{equation}
For simplicity, we can make the following choice for the incoming momenta:
\begin{equation}
|p_1+p_2|=|p_3+p_4|\equiv |p|,\,\,\,\,\vec{p}_1=-\vec{p}_2,\,\,\,\,\vec{p}_3=-\vec{p}_4;
\label{6-point-choice}
\end{equation}
thus, the amplitude in Eq.\eqref{6-point-cubic} is roughly given by
\begin{equation}
\mathcal{M}_6\sim\frac{\lambda^{12}}{(2!)^2}\frac{e^{-\frac{5}{3}(2(1)^2+2^2)p^2/M_s^2}}{p^{6}}=\frac{\lambda^{12}}{(2!)^2}\frac{e^{-10p^2/M_s^2}}{p^6},\label{6-point-rough}
\end{equation}
where we have neglected the terms such as $(p_3^4)^2$ and $2p_3^4p$ as $2p^2>(p_3^4)^2,\,2p_3^4p$, and this approximation becomes even more justified for a very large number of external legs, i.e. when $N\gg1$; see below. By adding two extra external legs, $p_7$ and $p_8$, and making similar choices as in Eq.\eqref{6-point-choice} and neglecting the cross-terms, one can see that the $8$-point scattering amplitude will behave as
\begin{equation}
\mathcal{M}_8\sim\frac{\lambda^{18}}{(3!)^2}\frac{e^{-\frac{5}{3}(2(1^2+2^2)+3^2)p^2/M_s^2}}{p^{10}}=\frac{\lambda^{18}}{(3!)^2}\frac{e^{-\frac{95}{3}p^2/M_s^2}}{p^{10}}.\label{8-point-rough}
\end{equation}
By inspecting Eqs.~(\ref{6-point-rough},~\ref{8-point-rough}), it is clear that by increasing the number of external legs, the scattering amplitude becomes even more exponentially suppressed. We can now easily find the expression for an $N$-point scattering amplitude, which will be roughly given by\footnote{The formula in Eq.\eqref{N-point-rough} is valid for $N$ even, but the analog formula for $N$ odd can be easily derived. However, in the large $N$ limit the results are the same and do not depend on the parity of the number of legs.}
\begin{equation}
\begin{array}{rl}
\mathcal{M}_N=& \displaystyle V_{\rm dress}(p_1,p_2,p_1+p_2)\prod\limits_{i=2}^{N-2}\Pi_{\rm dress}(p_1+\cdots+p_i)\\
&\displaystyle \times \prod_{j=2}^{N-3}V_{\rm dress}(p_1+\cdots+p_j,\,p_{j+1},\,p_1+\cdots+p_{j+1})\\
&\displaystyle \times V_{\rm dress}(p_1+\cdots+p_{N-2},p_{N-1},p_N)\\
\xrightarrow{UV}&\displaystyle  \frac{\lambda^{3(N-2)}}{[(N-2)/2]!^2}\frac{e^{-\frac{10}{3}\left(\sum\limits_{l=1}^{\frac{N-2}{2}}l^2-\frac{1}{2}\left(\frac{N-2}{2}\right)^2\right)p^2/M_s^2}}{p^{2(N-3)}},
\end{array}
\label{N-point-rough}
\end{equation}
where now the conservation law of the $4$-momenta in Eq.\eqref{cons-law-6} and the choice in Eq.\eqref{6-point-choice} generalize to
\begin{equation}
p_1+p_2+\cdots +p_{N-2}=p_{N-1}+p_{N}\label{cons-law-N}
\end{equation}
and
\begin{equation}
|p_i+p_{i+1}|\equiv|p|,\,\,\,\,\vec{p}_i=-\vec{p}_{i+1},\,\,\,\, i=1,\dots,N-3;
\label{N-point-choice}
\end{equation}
and we have used the relation 
\begin{equation}
2j^2p^2\gg (p_{2j+1}^4)^2,\,2p^4_{2j+1}jp,\label{neglect-condition}
\end{equation}
to neglect terms like $(p_{2j+1}^4)^2$ and $2p^4_{2j+1}jp,$ as $j\gg1.$ Note that the second set of equalities in Eq.\eqref{N-point-choice} corresponds to the choice of the centre of mass frame for $N-2$ incoming particles; indeed, for two incoming particles we would only have $\vec{p}_1=-\vec{p}_2$ and recover the usual relation between the spatial part of the two incoming momenta in the case of a $4$-point scattering amplitude.

The numeric series in Eq.\eqref{N-point-rough} can be summed up and in the limit $N\gg 1$ reads:
\begin{equation}
\sum\limits_{l=1}^{\frac{N-2}{2}}l^2=\frac{N(N-1)(N-2)}{12}\,\xrightarrow{N\gg1}\, N^3;\label{natural-squared}
\end{equation}
therefore, for a large number of interacting particles the $N$-point amplitude in Eq.\eqref{N-point-rough} shows the following behavior:
\begin{equation}
\mathcal{M}_N\sim\lambda'\frac{e^{-N^3p^2/M_s^2}}{p^{2(N-3)}}=\lambda'\frac{e^{-p^2/M_{\rm eff}^2}}{p^{2(N-3)}},\label{N-point-rough2}
\end{equation}
where we have defined $\lambda':=\lambda^{3(N-2)}/[(N-2)/2]!^2$ and in the last step we have introduced the {\it effective scale}
\begin{equation}
M_{\rm eff} \sim \frac{M_s}{N^{{3}/{2}}}.\label{eff-scatt-n=1}
\end{equation}
Hence, from Eqs.(\ref{N-point-rough2},~\ref{eff-scatt-n=1}) we have obtained that by increasing the number of external legs, or in other words the number of interacting particles, the scattering amplitude becomes more exponentially suppressed. This feature can be understood as follows:  there is a {\it transmutation} of scale under which the fundamental scale of nonlocality $M_s$ {\it shifts} towards lower energies, i.e. 
$M_{\rm eff}\ll M_s$ when $N\gg1$. In this process, the nonlocal length and time scales can be made much larger than the original scale of nonlocality, i.e. $M_{s}^{-1} \longrightarrow N^{3/2}M_{s}^{-1}$, therefore its affect can be felt in the IR. This phenomena of transmuting the scale from UV to IR has been shown in the fuzz ball construction in string theory setup to resolve blackhole singularity and horizon~\cite{Mathur:2005zp}.

The above calculations are performed for the form-factor $e^{-\Box/M_s^2},$ i.e. with $n=1.$ We can  generalize straightforwardly the previous results to generic powers $n$ of the d'Alembertian, such as $e^{(-\Box/M_s^2)^n}:$
\begin{equation}
\mathcal{M}_N\sim\lambda'\frac{e^{-\frac{10}{3}\left(\sum\limits_{l=1}^{\frac{N-2}{2}}l^{2n}\right)p^{2n}/M_s^{2n}}}{p^{2(N-3)}}.\label{N-point-rough n}
\end{equation}
The numeric series in Eq.\eqref{N-point-rough n} can be expressed in terms of the {\it Faulhaber formula} which is given by~\cite{numbers}
\begin{equation}
\!\sum\limits_{l=1}^{\frac{N-2}{2}}l^{2n}=\frac{1}{2n+1}\sum\limits_{i=0}^{2n}(-1)^i{2n+1\choose i}B_i\cdot \left(\frac{N}{2}-1\right)^{2n+1-i}\!,
\label{faulhaber}
\end{equation}
where $B_i$ is the so called Bernoulli number. The expression in Eq.\eqref{faulhaber} seems rather complicated, but fortunately we are only interested in the limit $N\gg1,$ which gives
\begin{equation}
\sum\limits_{l=1}^{\frac{N-2}{2}}l^{2n}\,\xrightarrow{N\gg1}\, N^{2n+1}.
\label{faulhaber-limit}
\end{equation}
Hence, the $N$-point scattering amplitude for generic powers $n$ of the d'Alembertian will behave as
\begin{equation}
\mathcal{M}_N\sim\lambda'\frac{e^{-(N^{2n+1})p^{2n}/M_s^{2n}}}{p^{2(N-3)}}=\lambda'\frac{e^{-p^{2n}/M_{\rm eff}^{2n}}}{p^{2(N-3)}}\label{N-point-rough n 2},
\end{equation}
where in this more general case the effective nonlocal scale is defined as:
\begin{equation}
M_{\rm eff}\sim \frac{M_s}{N^{\frac{2n+1}{2n}}}=\frac{M_s}{N^{1+{1}/{2n}}}\label{eff-scatt-n}.
\end{equation}
%

%%%%%%%%%%%%%%%%%%%%%%%%%%%%%%%%
\subsection{Zero external momenta}

 We now wish to ask a similar question but for a different kind of amplitude, with {\it zero} ingoing and outgoing external momenta. As done before, let us start when $n=1$, and consider a tree-level diagram as the one in Fig. $1$, but with a different choice of the external momenta. For this kind of diagram (with no loops) we can not set all single momentum equal to zero, otherwise we would not get any exponential contributions, but we will consider the following choice:
\begin{equation}
p_1+\cdots +p_{N-2}=p_{N-1}+p_{N},
\end{equation}
\begin{equation}
|p_i+p_{i+1}|\equiv |p|,\,\,\,\,\,i=1,\dots,N-3,
\end{equation}
and 
\begin{equation}
\begin{array}{ll}
\displaystyle p_{N}=-p_{N-1},&\\
\displaystyle p_i+p_{i+1}=-(p_{i+2}+p_{i+3}),\,\,\,\,i=1,\dots, N-5;&
\end{array} 
\label{N-point-choice-zero}
\end{equation}
with on-shell conditions $p_i^2=0.$ 

For the above choice of momentum distribution in Eq.\eqref{N-point-choice-zero}, the IR divergences from the denominators may appear. However, they can be cured as in the standard local field theory where non locality does not play any role. Indeed they are just related to the fact that we are working with a massless scalar field. Anyway, we are interested in the regime where nonlocality in the propagator becomes important, and want to understand the role played by the exponential form-factors. 
\begin{figure}[t]
	\begin{center}
		\includegraphics[scale=0.52]{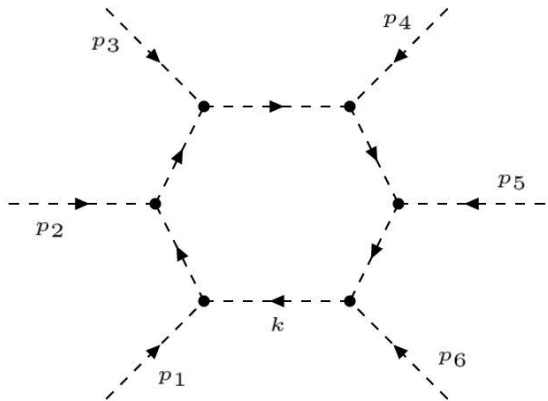}
		\caption{Ring diagram: $1$-loop $6$-point amplitude for scalar cubic interaction. The blobs correspond to dressed vertices.}
	\end{center}\label{Fig_2}
\end{figure}
In fact, in this regime the tree-level $N$-point amplitude, in the limit $N\gg 1,$ will be given by
\begin{equation}
\mathcal{M}_N\xrightarrow{N\gg1}\lambda^{3(N-2)}e^{-Np^{2}/M_s^{2}}=\lambda^{3(N-2)}e^{-p^{2}/M_{\rm eff}^{2}},\label{N-point-zero-1}
\end{equation}
where we have introduced the {\it effective scale}
\begin{equation}
M_{\rm eff}\sim \frac{M_s}{\sqrt{N}}.\label{eff-scale-zero-1}
\end{equation}
Hence,  in the case of zero ingoing and outgoing external momenta the total amplitude becomes more suppressed for an increasing number of interacting particles, but by comparing to the previous case, see Eq.\eqref{eff-scatt-n=1}, the scaling is different.

Furthermore, we can show that similar transmutation of the scale manifest also for different amplitudes, as for the $1$-loop diagram of the kind in 
Fig.$2$, known as the {\it ring diagram}. In this case, given $N$ external legs we have $N$ vertices and $N$ internal propagators. Since we have a loop in the diagram, we can now set all individual external momenta to zero, and we can show that an $N$-point amplitude as the one in Fig. $2$, with $p_i=0,$ $i=1,\dots,N,$ reads:
\begin{equation}
\begin{array}{rl}
\mathcal{M}_{{\rm ring},\,N}\sim& \displaystyle \lambda^{3N}\int \frac{d^4k}{(2\pi)^4} \frac{e^{-Nk^2/M_s^2}}{k^{2N}}\\
=&\displaystyle \lambda^{3N}\int \frac{d^4k}{(2\pi)^4} \frac{e^{-k^2/M_{\rm eff}^2}}{k^{2N}},\label{ampl-1-loop}
\end{array}
\end{equation}
where the effective scale $M_{\rm eff}$ coincides with the one in Eq.\eqref{eff-scale-zero-1}.
\begin{figure}[t]
	\begin{center}
		\includegraphics[scale=0.50]{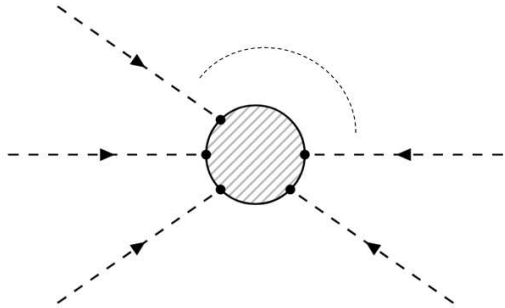}
		\caption{The dashed blob represents the region in which the nonlocal interaction takes place. The larger is the number of interacting particles, the larger will be the nonlocal region in coordinate space and time.}
	\end{center}\label{Fig_3}
\end{figure}
So far we have only considered the scaling properties for $n=1,$ but we can straightforwardly  generalize the above results to generic 
$\Box^{n}$. We can show that for an $N$-point amplitude with zero external momenta, the scale of nonlocality will transmute to
\begin{equation}
M_{\rm eff}\sim \frac{M_s}{N^{1/2n}}.\label{eff-scale-zero-n}
\end{equation}
% 

%%%%%%%%%%%%%%%%%%%%%%%%%%%%%%%%%%%%%%%%%%%%%%%%%%%%%%%%%%%%%%%%
\section{Scalar toy-model for infinite derivative gravity}\label{toy-model-gravity}

We now wish to consider a slightly more interesting scenario of a scalar toy-model, which can mimic the graviton self-interaction in ghost-free infinite derivative gravity (IDG), up to cubic order $\mathcal{O}(h^{3}),$ where $h$ is the trace $h=\eta^{\mu\nu}h_{\mu\nu}$ of the graviton perturbation around the Minkowski background, which is now mimed by the scalar field $\phi(x).$  
Such a model was first studied in Refs. \cite{Talaganis:2014ida,Talaganis:2016ovm} and the corresponding Euclidean action reads~\cite{Talaganis:2014ida}~\footnote{In Ref. \cite{Talaganis:2017dqy}, the authors computed $N$-point amplitudes for a simpler version of this action. However, the authors only considered the power $n=1$ and the case of zero external momenta. Moreover, the choice they made for the momenta $p_i$ seems to be not physically sensible, and it is different from ours in Eq.\eqref{N-point-choice-zero}.}
\begin{equation}
\begin{array}{rl}
S=& \displaystyle \int d^4x \left( \frac{1}{2} \phi(x)e^{\left(-\Box/M_s^2\right)^n}\Box\phi(x)\right.\\
& +\displaystyle\frac{\lambda}{4}\phi(x)\partial_{\mu}\phi(x)\partial^{\mu}\phi(x)\\
&\displaystyle+\frac{\lambda}{4}\phi(x)\Box\phi(x)e^{\left(-\Box/M_s^2\right)^n}\phi(x)  \\
&\displaystyle\left.-\frac{\lambda}{4}\phi(x)\partial_{\mu}\phi(x)e^{\left(-\Box/M_s^2\right)^n}\partial^{\mu}\phi(x)\right).
\end{array}\label{action-toy}
\end{equation}
The above action exhibits the following scaling symmetry:  $\phi\longrightarrow (1+\epsilon)\phi+\epsilon$ \cite{Talaganis:2014ida}.
The Euclidean bare propagator for this action is the same as the one in Eq.\eqref{euclid-propag}, while the bare vertex is not a constant, but it is given by \cite{Talaganis:2014ida,Talaganis:2016ovm}
\begin{equation}
\!\!\begin{array}{rl}
V(k_1,k_2,k_3)=&\displaystyle\frac{\lambda}{4}\left(k_1^2+k_2^2+k_3^2\right)\left(e^{k^{2n}_1/M_s^{2n}}+e^{k^{2n}_2/M_s^{2n}}\right.\\
&\displaystyle\,\,\,\,\,\,\,\,\,\,\,\,\,\,\,\,\,\,\,\,\,\,\,\,\,\,\,\,\,\,\,\,\,\,\,\,\,\,\,\,\,\,\,\,\,\,\,\left.+e^{k^{2n}_3/M_s^{2n}}-1\right).
\end{array}
\label{vertex-toy}
\end{equation}
Even though the propagator is exponentially suppressed, the vertex function is exponentially enhanced.
We will make explicit computations for the power $n=1$, and then generalize to any power of $\Box^{n}$.

%%%%%%%%%%%%%%%%%%%%%%%%%%%%%%%%%%

\subsubsection{Dressing the propagators}

Unlike the case of the cubic interaction in Section \ref{scal-cub} (see Eq.\eqref{cubic-action}), in the case of the above action in Eq.\eqref{action-toy}, the UV behavior of the propagator is slightly modified by loop quantum corrections, as shown in Refs.~\cite{Talaganis:2014ida,Talaganis:2016ovm}. First of all, the self-energy at $1$-loop for the action in Eq.\eqref{action-toy} is given by
\begin{equation}
\Sigma^{(1)}(p)= \displaystyle \int \frac{d^4k}{(2\pi)^4}\frac{V^2(p,k,k-p)}{k^2(k-p)^2}e^{-\frac{k^2}{M_s^2}}e^{-\frac{(k-p)^2}{M_s^2}}.
\end{equation}
The above integral can be exactly computed and reads~\cite{Talaganis:2014ida}
\begin{equation}
\!\!\!\begin{array}{ll}
\displaystyle\Sigma^{(1)}(p)= \displaystyle \frac{\lambda^2}{16\pi^2}\left\lbrace \frac{p^4}{8} \left({\rm log}\left(\frac{p^2}{4\pi M_s^2}\right)+\gamma_E-2\right)  \right.&\\[3mm]
\,\,\displaystyle+\frac{e^{-p^2/M_s^2}M^2_s}{32p^2}\left[2M_s^2p^2e^{p^2/M_s^2}\left(e^{2p^2/M_s^2}-1\right){\rm Ei}\left(-\frac{p^2}{M_s^2}\right)\right.&\\[3mm]
\,\displaystyle-\left(e^{p^2/M_s^2}-1\right)\left(M_s^2p^2e^{p^2/M_s^2}\left(e^{p^2/M_s^2}-1\right){\rm Ei}\left(-\frac{p^2}{2M_s^2}\right)\right.&\\[3mm]
\,\,\displaystyle +\left(e^{\frac{3}{2}p^2/M_s^2}-e^{p^2/2M_s^2}\right)\left(2p^4+5M_s^2p^2+4M_s^4\right)&\\[3mm]
\,\,\,\,\,\,\,\,\,\,\,\,\displaystyle +2e^{p^2/M_s^2}\left(7p^4+7M_s^2p^2+2M_s^2\right)&\\[3mm]
\,\,\,\,\,\,\,\,\,\,\displaystyle \left.\left.\left.-2\left(p^4+3M_s^2p^2+2M_s^4\right)\right)\right]\right\rbrace, &
\label{1-loop self-energy-deriv}
\end{array}
\end{equation}
where $\gamma_E=0.57721\dots$ is the Euler-Mascheroni constant. Although the above expression is very complicated, in the UV regime the behavior of the $1$-loop self-energy turns out to be simpler and is given by
\begin{equation}
\Sigma^{(1)}(p)
\xrightarrow{UV}\displaystyle \lambda^2 e^{\frac{3}{2}p^2/M_s^2},
\label{1-loop self-energy-deriv-uv}
\end{equation}
where we have neglected numerical factors. With the help of Eqs.\eqref{1-loop self-energy-deriv},\eqref{1-loop self-energy-deriv-uv}, we can now compute the dressed propagator, whose UV behavior is given by
\begin{equation}
\Pi_{\rm dress}(p)~~\xrightarrow{UV}~~\lambda^{-2}e^{-\frac{3}{2}p^2/M_s^2},
\label{uv-dressed-deriv}
\end{equation}
which turns out to be even more exponentially suppressed than the bare one, as we now have the factor $3/2$ in the exponent as compared to 
Eq.\eqref{euclid-propag}.

%%%%%%%%%%%%%%%%%%%%%%%%%%%%%%%%%%

\subsubsection{Dressing the vertices}

In Ref.~\cite{Talaganis:2016ovm}, it was shown that eventhough a bare vertex is exponentially enhanced the dressed vertex can be exponentially suppressed, provided higher order quantum loops, built with dressed propagators, are taken into account. In fact, in the UV regime the dressed vertex has the following form~\cite{Talaganis:2016ovm}:
\begin{equation}
V_{\rm dress}^{(l)}(k_1,k_2,k_3)~~\xrightarrow{UV}~~\lambda^3 e^{\alpha^{(l)}\frac{k_1^2}{M_s^2}+\beta^{(l)}\frac{k_2^2}{M_s^2}+\gamma^{(l)}\frac{k_3^2}{M_s^2}},
\end{equation}
where $l$ is the number of loops, or in other words, the loop order. If $l\geq 4$, the dressed vertex becomes exponentially suppressed; for instance, if $l=4$ we obtain \cite{Talaganis:2014ida,Talaganis:2016ovm}:
\begin{equation}
\alpha^{(4)}=\beta^{(4)}=\gamma^{(4)}=-\frac{11}{27},
\end{equation}
so that the dressed vertex takes the following form in the UV:
\begin{equation}
V_{\rm dress}^{(4)}(k_1,k_2,k_3)~~\xrightarrow{UV}~~\lambda^3e^{-\frac{11}{27}\left(\frac{k_1^2}{M_s^2}+\frac{k_2^2}{M_s^2}+\frac{k_3^2}{M_s^2}\right)}.\label{dress-vertex-deriv}
\end{equation}
%

%%%%%%%%%%%%%%%%%%%%%%%%%%%%%%%%%%%%%%%%%%%%%

\subsection{$N$-point scattering amplitude}

By assuming first the simple case, $n=1,$ we will compute the $4$-point scattering amplitude, with momenta $p_1+p_2=p_3+p_4\equiv p,$ where we use both dressed propagators and vertices, the latter at the loop order $l=4$:
\begin{equation}
\begin{array}{rl}
\mathcal{M}_4=&\displaystyle V_{\rm dress}^{(4)}(p_1,p_2,p)\Pi_{\rm dress}(p)V_{\rm dress}^{(4)}(p,p_3,p_4)\\[3mm]
\,\,\xrightarrow{UV}& \displaystyle \lambda^4 e^{-\left((2)\frac{11}{27}+\frac{3}{2}\right)p^2/M_s^2}\sim\lambda^4 e^{-\frac{125}{54}p^2/M_s^2},
\end{array}
\label{4point-deriv}
\end{equation}
which turns out to be exponentially suppressed in Euclidean signature, where we have used the on-shell condition $p_i^2=0.$  
Let us now consider the dressed $N$-point amplitude, with $N>4:$
\begin{equation}
\!\!\!\begin{array}{ll}
\mathcal{M}_N=\displaystyle V_{\rm dress}^{(4)}(p_1,p_2,p_1+p_2)\prod\limits_{i=2}^{N-2}\Pi_{\rm dress}(p_1+\cdots+p_i)&\\
\,\,\,\,\,\,\times \displaystyle \prod_{j=2}^{N-3}V_{\rm dress}^{(4)}(p_1+\cdots+p_j,\,p_{j+1},\,p_1+\cdots+p_j+p_{j+1})&\\
\,\,\,\,\,\,\times \displaystyle V_{\rm dress}^{(4)}(p_1+\cdots+p_{N-2},p_{N-1},p_{N});&
\end{array}
\label{Npoint-deriv}
\end{equation}
when $N=4$ the product in the second line is just one, and we recover the result in Eq.\eqref{4point-deriv}. 
By imposing the on-shell conditions $p_i^2=0$, and making the choices for momenta as in Eqs.~(\ref{N-point-choice},~\ref{neglect-condition}), the UV behavior of the $N$-point amplitude in Eq.\eqref{Npoint-deriv} reads:
\begin{equation}
\mathcal{M}_N~~\xrightarrow{UV}~~\lambda^{N}e^{-\frac{125}{27}\left(\sum\limits_{l=1}^{\frac{N-2}{2}}l^2-\frac{1}{2}\left(\frac{N-2}{2}\right)^2\right)p^2/M_s^2}.
\label{Npoint-deriv2}
\end{equation}
By assuming the limit $N\gg 1$, we obtain
\begin{equation}
\mathcal{M}_N~~\xrightarrow{N\gg1}~~\lambda^{N} e^{-N^3p^2/M_s^2}=\lambda^{N} e^{-p^2/M_{\rm eff}^2}.
\label{Npoint-deriv2-behavior}
\end{equation}
where we have introduced the {\it effective scale} 
\begin{equation}
M_{\rm eff}\sim \frac{M_s}{N^{{3}/{2}}},\label{scaling-2}
\end{equation}
which coincides with the scaling in Eq.\eqref{eff-scatt-n=1}.

So far we have only considered $n=1$ case, the calculations are complicated for generic powers of $n$ of the d'Alembertian. However, we can still understand the problem by observing that the UV behaviors of dressed propagators and vertices are proportional to $e^{-c\left(p^2/M_s^2\right)^n},$ with some positive numerical factor, $c>0.$  Thus, we can generalize our results to generic powers of $n,$ and show that the scaling still coincides with the one obtained for the action in Eq.\eqref{cubic-action} (see Eq.\eqref{eff-scatt-n}):
\begin{equation}
M_{\rm eff}\sim \frac{M_s}{N^{\frac{2n+1}{2n}}}.\label{eff-scale-scatt-n-deriv}
\end{equation}
% 

%%%%%%%%%%%%%%%%%%%%%%%%%%%%%%%%%%%%%%%%

\subsection{Zero external momenta}

We now wish to compute the $N$-point amplitudes with zero ingoing and outgoing external momenta. In the UV when $N\gg1$, the amplitude follows the same behavior as the one for the scalar action in Section \ref{scal-cub}. Indeed, by dressing both internal propagators and vertices as done in Eqs.(\ref{uv-dressed-deriv},~\ref{dress-vertex-deriv}), we can  show that we get a similar result as in Eqs.(\ref{N-point-zero-1},~\ref{ampl-1-loop}). For instance, the ring diagram in Fig. $2$ with zero external momenta reads:
\begin{equation}
\begin{array}{rl}
\mathcal{M}_{{\rm ring},\,N}\sim& \displaystyle \lambda^{N}\int \frac{d^4k}{(2\pi)^4} e^{-\frac{125}{54}Nk^{2n}/M_s^{2n}}\\
\xrightarrow{N\gg1}&\displaystyle \lambda^{N}\int \frac{d^4k}{(2\pi)^4} e^{-k^{2n}/M_{\rm eff}^{2n}}.\label{ampl-1-loop-deriv}
\end{array}
\end{equation}
Thus, the scale of nonlocality now transmutes as in our previous case, see Eq.\eqref{eff-scale-zero-n}):
\begin{equation}
M_{\rm eff}\sim \frac{M_s}{N^{\frac{1}{2n}}}.\label{eff-scale-zero-n-deriv}
\end{equation}
% 

%%%%%%%%%%%%%%%%%%%%%%%%%%%%%%%%%%%%%%%%%%%%%%%%%%%%%%%%%%%

\section{Discussions and Conclusions}\label{concl}

In this paper we have computed the $N$-point amplitudes in the context of ghost-free infinite derivative scalar field theories. We have worked with a massless scalar field and studied two toy models in Eqs.(\ref{cubic-action},~\ref{action-toy}). In particular, we were interested in the limit in which the number of particles, $N,$ could be very large and we have shown that the scale of nonlocality, $M_s,$ transmutes to a lower value in the IR depending on $N$, and on the form of the entire function. Although, $M_s$ represents a physical cut-off and beyond which it is hard to probe the nature of physics, but with a large number of interacting particles/quanta the nonlocal regime becomes more accessible in the IR.

The above computations also provide a tangible support to IDG theories, where the scale of nonlocality may be a dynamical quantity and can be modified in presence of a large number of gravitons interacting nonlocally, as argued from completely different point of view - by arguing that the entropy of a gravitationally bound system would follow the {\it Area}-law~\cite{Koshelev:2017bxd}. This result has already played a key role in constructing non-singular compact objects~\cite{Koshelev:2017bxd,Buoninfante:2018rlq}. However, explicit scattering amplitude computations for $N$-point amplitudes in the context of nonlocal gravitational interaction are still lacking and will be the subject of future works.

\acknowledgements  The authors are thankful to Valeri Frolov, Nobuchika Okada, Masahide Yamaguchi, Joao Marto  and Alexey Koshelev.
AM's research is financially supported by Netherlands Organisation for Scientific Research (NWO) grant number 680-91-119. AG is supported by LNF-INFN \& University Roma Tre for logistics and support and thanks GGI for various interesting discussions during the JH workshop (October 2018).

%%%%%%%%%%%%%%%%%%%%%%%%%%%%%%%%%%%%%%%%%%%%%%%%%%%%%%%%%%%%%%%%%%%%%%%%%%%%%%%%%%%%%%%%%%%

\end{document}